\documentclass{jltp}
\usepackage{jltp}  
\def\gapx{\lower 2pt
\hbox{$\buildrel>\over{\scriptstyle{\sim}}$}}
\def\lapx{\lower 2pt \hbox{$\buildrel<\over{\scriptstyle{\sim}}$}}
\def\4he{$^4$He}

\def\br{{\bf r}}

\def\beq{\begin{equation}}
\def\eeq{\end{equation}}
\def\bea{\begin{eqnarray}}
\def\eea{\end{eqnarray}}

\def\bv{{\bf v}}

\title{Two-Fluid Hydrodynamics in Trapped Bose Gases and in Superfluid 
Helium}

\author{Allan Griffin and Tetsuro Nikuni}

\address{Department of Physics, University of Toronto\\Toronto,
Ontario M5S 1A7, Canada}

\runninghead{A. Griffin and T. Nikuni}{Two-Fluid Hydrodynamics}

\begin{document}

\begin{abstract}
A review is given of recent theoretical work on the superfluid dynamics of 
trapped Bose gases at finite temperatures, where there is a significant 
fraction of non-condensate atoms.    One can now reach large enough 
densities and collision cross-sections needed to probe the collective 
modes in the collision-dominated hydrodynamic region where the gas 
exhibits characteristic superfluid behavior involving the relative motions 
of the condensate and non-condensate components.  The precise analogue of 
the Landau-Khalatnikov two-fluid hydrodynamic equations was recently 
derived from trapped Bose gases, starting from a generalized 
Gross-Pitaevskii equation for the condensate macroscopic wavefunction and 
a kinetic equation for the non-condensate atoms.  

\end{abstract}

\maketitle

\section{INTRODUCTION}
\label{sec:Introduction}

Superfluid behavior is the most striking property of liquid  $^4$He.  We 
recall that spectacular experiments of Kapitza as well as Allen and 
Misener in 1938 first showed that liquid $^4$He below the transition 
temperature of 2.19K could exhibit flow through thin channels without any 
viscosity.  Since then, a dominant theme of research on liquid $^4$He has 
been to understand the origin of this superfluidity and to work out how it 
modifies the dynamical response functions and excitations of liquid 
$^4$He. 

By the early 1960's, a successful field theoretic formalism was 
developed\cite{Hohmar,review}, with the broken-symmetry expectation value 
of the quantum field operator $\Phi(\br, t) = \langle\hat\psi({\bf r}, 
t)\rangle$ playing the role of the order parameter of the Bose 
superfluid.  In this theory, Bose-condensation was the underlying source 
of {\em all} the unique properties of superfluid $^4$He (equivalence of 
density fluctuations with elementary excitations, two-fluid hydrodynamical 
behavior at low frequencies, superfluid flow, quantized vortices, etc). In 
striking contrast to the atomic Bose gases, the explicit role of the Bose 
condensate is somewhat hidden in superfluid $^4$He.  Indeed, this fact has 
encouraged the development of theories for superfluid $^4$He based on 
variational groundstate wavefunctions (beginning with Feynman in the 
mid-1950's) without any explicit reference to the underlying role of the 
macroscopic wavefunction describing the condensate.   While very useful 
for computational purposes, these approaches have given little insight 
into the fact that there is a new phase in liquid $^4$He which exhibits 
superfluidity.  For this reason,  they do not appear to give much hope for 
developing a general theory of the superfluid dynamics of Bose fluids, 
going from dilute gases to liquid $^4$He.

Such a unified description of this kind has taken on a new importance as a 
result of spectacular advances that have occured in BEC research in the 
last year.  These discoveries have started to focus research interest in 
atomic gases on issues related to superfluidity.  The recent creation of 
quantized vortices (at JILA in Boulder\cite{Madison} and the ENS in 
Paris\cite{Matand}) have suddenly made the BEC community aware of rotating 
traps, mutual friction, vortex arrays, etc, ideas long familiar to the 
superfluid $^4$He community.  In addition, the successful creation of a 
BEC in $^{85}$Rb gas at JILA\cite{Cornish} by working close to  a Feshbach 
resonance has given us a superfluid Bose gas where the s-wave scattering 
length $a$ can be enormous $(a> 10^3 \stackrel{\circ}{\rm A})$.  Already 
this has moved the value of the gas parameter $(na^3)$ from the range 
$\sim 10^{-5}$ (typical for recent BEC experiments\cite{Ingstrwie} using 
$^{87}$Rb and $^{23}$Na atoms) to a respectable $10^{-2}$.  Thus atomic 
condensates are ``starting'' to overlap on a strongly interacting Bose 
liquid like $^4$He (where $na^3 \sim 1$, with $a$ now being the hard core 
diameter).

More generally, the BEC community is increasingly using the language and 
ideas developed in condensed matter and quantum liquid research.  
Arguments based on topological arguments are used (untwisting of order 
parameters\cite{Matandhal}, for example). Recent Bragg experiments using 
second-order Raman scattering of light have allowed one to 
measure\cite{Stamkurn} the dynamic structure factor $S(\bf q, \omega)$ of 
trapped Bose gases, a quantity which has played a central role in our 
understanding of superfluid $^4$He for almost 50 years.\cite{review} In 
the last few years, much of the theoretical analysis of atomic condensates 
has come from condensed matter theorists who have worked on superfluid 
liquids (such as Baym, Fetter, Ho, Leggett, Pethick, Pitaevskii and  
Stringari).  We  hope that the superfluid $^4$He experimental community 
will be stimulated by the new BEC child in their midst and, in particular, 
come up with new ideas about how to probe the Bose-condensed nature of 
superfluid $^4$He.  

As a contribution to this emerging synthesis, this article reviews recent 
work\cite{Zarnikgrif,Nikuni99} on deriving (from an approximate but still 
microscopic model) the two-fluid hydrodynamic equations for a trapped Bose 
gas at finite temperatures.  The equivalent two-fluid equations were first 
derived phenomenologically by Landau in 1941, initially without any 
reference to a Bose condensate.   Generalized to include damping from 
transport coefficients, these coupled equations for the superfluid and 
normal fluid components form the basis of our understanding of all 
``superfluid'' behavior in liquid $^4$He.\cite{Khalatnikov}  This 
two-fluid description was later proven to be a consequence of a Bose 
broken-symmetry by Bogoliubov\cite{Bogg} in 1963, with the superfluid 
velocity ${\bf v}_s(\br, t)$ always being proportional to the gradient of 
the phase of $\Phi(\br, t)$.
 
Of course, the two-fluid hydrodynamic equations describe local equilibrium 
and require that there be enough collisions to achieve this 
$(\omega\tau\ll 1,$ where $\tau$ is the relaxation time needed to reach 
local equilibrium).   This is easy to achieve in a liquid (including 
liquid $^4$He) and that is why, historically, classical fluid 
hydrodynamics was understood a long time before a microscopic treatment 
was available.  Using a kinetic equation for interacting atoms, it was 
Boltzmann who first ($\sim$ 1885) discussed the precise set of conditions 
required for this hydrodynamic description to be valid.  Up to the 
present, most experiments on the collective modes of  Bose gases have 
probed the opposite ``collisionless region'' $(\omega\tau\gg 1).$  However 
one expects that the two-fluid hydrodynamic region will be increasingly 
studied in the next few years, taking advantage of the fact that one can 
now produce high densities of atoms (currently, $N\sim 10^6-10^7$ atoms) 
and also we can enormously increase the atomic collision cross-section 
$(\sigma = 8\pi a^2)$ by working near a Feshbach resonance (as in recent 
work\cite{Cornish} on $^{85}$Rb). 

As references on BEC research, we recommend the 1998 Varenna Summer School 
Lectures\cite{Ingstrwie} as well as the excellent review\cite{Dalgiorpit} 
by the Trento theory group.  For a detailed treatment of topics discussed 
here, see Refs.\onlinecite{Grifsum,Zarnikgrif}.
\vskip 1.5 true cm
\section{DYNAMICS OF A PURE CONDENSATE}
\label{sec:Dynamics}

The crucial idea behind the field-theoretic description of Bose 
condensation developed 40 years ago\cite{Hohmar,review} is to isolate the 
condensate ``degree of freedom.''  Thus the quantum field operator is 
decomposed as $\hat\psi(\br) = \langle\hat\psi(\br)\rangle + 
\tilde\psi(\br),$
where  $\Phi(\br) \equiv \langle\hat\psi(\br)\rangle$ describes the Bose 
condensate (which is treated as a classical field).  Here 
$\tilde\psi(\br)$ is the non-condensate component of the quantum field 
operator and satisfies Bose commutation relations.  This formalism (first 
developed in a systematic way by Beliaev in 1958) allows one to neatly 
isolate the condensate.  The many-body theory is then developed to study 
the dynamics of the non-condensate fields $\tilde\psi$ and 
$\tilde\psi^\dagger$, with the condensate playing the role of the 
``vacuum.''  A crucial feature is that $\Phi(\br, t)$ is complex,

\begin{equation}
\Phi(\br, t) = \sqrt{n_c({\bf r}, t)}e^{i\theta({\bf r}, 
t)},\label{dynamics-eq2} 
\end{equation}
with the identification (this is justified later) of the superfluid 
velocity $m{\bf v}_s(\br)$ $= \hbar{\mbox{\boldmath$\nabla$}}\theta(\br, 
t).$  At $T=0$ in a dilute Bose gas, one can assume that all the atoms are 
in the condensate $\Phi (\br, t),$ which satisfies the famous 
Gross-Pitaevskii time-dependent equation of motion\cite{Dalgiorpit},
\begin{equation}
i\hbar {\partial\Phi(\br, t)\over\partial t} = 
\left[-{\hbar^2\nabla^2\over 2m}+V_{ex} ({\bf 
r})+gn_c({\br},t)\right]\Phi({\br }, t). \label{dynamics-eq3}
\end{equation}
Here $V_{ex} (\br)$ is the parabolic trap potential (a magnetic trap 
acting on the spin of an hyperfine atomic level) and $g n_c(\br, t)$ is 
the Hartree field produced by the condensate atoms.  At the very low 
temperatures of interest in BEC studies, the atoms have very low energy 
and one  can use the $s$-wave scattering length approximation.  In this 
case, the effective interatomic interaction is given by $v(\br - 
\br^\prime) = g\delta(\br-\br^\prime),$ where $g = 4\pi a {\hbar^2}/m.$  
One should not think of (\ref{dynamics-eq3}) as simply a Schrodinger 
equation for a ``single-particle'' wavefunction, since $\Phi (\br, t)$ is 
an order-parameter which is well defined even at finite temperatures (see 
later).  As we all have learned over the last 5 years, the $T=0$ GP 
equation in (\ref{dynamics-eq3}) contains a huge amount of physics and its 
solutions have dominated BEC research up to 
now.\cite{Ingstrwie,Dalgiorpit}  Many examples are discussed in the 
invited and contributed papers at this QFS Conference.

The stationary solutions that the GP equation $\Phi_0(\br, t) = 
\Phi_0(\br)e^{-i\mu_0t/\hbar}$ satisfy
\beq \left[-{\hbar^2\nabla^2\over 2m} +V_{ex}(\br) + 
g|\Phi_0(\br)|^2\right]\Phi_0(\br)=\mu_0\Phi_0(\br).\label{dynamics-eq4}\eeq

This can exhibit ground state solutions corresponding to vortices, in 
which there is a time-independent condensate current given by $m{\bf 
v}_{s0}(\br) = \hbar{\mbox{\boldmath$\nabla$}}\theta_0(\br).$  In the 
absence of vortices (irrotational flow), the Thomas-Fermi (TF) approximate 
solution of (\ref{dynamics-eq4}) is very simple and leads to the famous 
parabolic condensate profile
\beq n_{c0}(\br) \equiv|\Phi_0(\br)|^2 ={1\over g}\left[\mu_0-{1\over 2} 
m\omega^2_0 r^2\right].\label{dynamics-eq5}\eeq
This profile is considerably wider than the ground state Gaussian 
wavefunction prediction for a non-interacting trapped gas.

It is very convenient to rewrite the GP equation (\ref{dynamics-eq3}) in 
terms of the condensate density $n_c(\br, t)$ and velocity $\bv_s(\br, t)$ 
local variables.  One finds
\bea {\partial n_c(\br, t)\over\partial t} &=& - 
\mbox{\boldmath$\nabla$}\cdot n_c(\br, t)\bv_s(\br, t)\nonumber\\
m\left({\partial\bv_s\over\partial t} +{1\over 
2}{\mbox{\boldmath$\nabla$}}\bv_s^2\right) &=& 
-{\mbox{\boldmath$\nabla$}}\mu_c(\br, t), \label{dynamics-eq6}
\eea
where the condensate chemical potential is
\beq \mu_c(\br, t) \equiv -{\hbar^2\nabla^2\sqrt{n_c}\over 
2m\sqrt{n_c}}+V_{ex}(\br) + gn_c(\br, t). \label{dynamics-eq7} \eeq
Written in this form, the condensate dynamics is ``hydrodynamic'' looking, 
even in the absence of collisions.  This is because the GP equation 
(\ref{dynamics-eq3}) describes a large number of atoms in the {\em same} 
single-particle quantum state.

Linearizing around $\Phi_0(\br)$, the two coupled equations in 
(\ref{dynamics-eq6}) can be solved for the collective oscillations of the 
condensate at $T=0$.  A standard simplification (first introduced by 
Stringari\cite{Sstringari}) is to neglect the ``quantum pressure'', the 
first term in (\ref{dynamics-eq7}).  Within this TF approximation, one 
obtains the Stringari ``wave equation'' for fluctuations in $\delta 
n_c(\br, t)$
\beq {\partial^2\delta n_c\over \partial t^2} = - 
\mbox{\boldmath$\nabla$}\cdot\left(n_{c0}{\partial\delta\bv_s\over\partial 
t}\right) = {g\over 
m}\mbox{\boldmath$\nabla$}\cdot\left(n_{c0}(\br)\mbox{\boldmath$\nabla$}\delta 
n_c\right).\label{dynamics-eq8}\eeq
We note that this wave equation could be equally well written in terms of 
fluctuations of the phase of $\Phi(\br, t)$, since 
$\partial\delta\theta/\partial t = -g\delta n_c$.  As an example of such 
condensate oscillations, we consider a uniform Bose gas, where $\delta 
n_c(\br, t)\propto e^{i({\bf k}\cdot{\bf r}-\omega t)}.$  This gives the 
famous Bogoliubov phonon excitations of the condensate (first derived in 
1947 by a different method)
\beq \omega = c_0k, \ \ \ \ \ c_0=\sqrt{gn_{c0}\over 
m}.\label{dynamics-eq9} \eeq
These phonons are physically unrelated to ordinary (hydrodynamic) sound 
waves in a normal fluid.  Solutions of (\ref{dynamics-eq8}) corresponding 
to breathing, dipole, quadrupole and surface oscillations of an axially 
symmetric trap are in excellent agreement with experiment (when $N\ \gapx\ 
10^4$ atoms).\cite{Dalgiorpit,Sstringari}

\section{TWO-FLUID SUPERFLUID HYDRODYNAMICS}
\label{sec:Superfluid}

It is straightforward to extend the preceding $T=0$ analysis to finite 
temperatures where there is a large fraction of atoms outside of the 
condensate described by $\Phi(\br, t).$  One then needs {\em two} 
equations of motion\cite{Zarnikgrif}:
(a)  A generalized GP equation for $\Phi(\br, t)$ which includes coupling 
to the non-condensate atoms. (b)  A Boltzmann kinetic equation for the 
single-particle distribution function $f({\bf p}, \br, t)$ describing the 
non-condensate atoms.

A simple microscopic model has been analyzed in some detail in 
Ref.\onlinecite{Zarnikgrif} which appears to contain the correct physics.  
By resticting oneself to finite temperatures, the important non-condensate 
atoms can be described by the simple particle-like spectrum
\bea
{\tilde\varepsilon}_p({\bf r}, t) &=& {p^2\over 2m} + V_{ex}({\bf 
r})+2g\left[n_c({\bf r}, t)+{\tilde n}({\bf r}, t)\right]\nonumber\\
&\equiv& {p^2\over 2m} + U({\bf r}, t).\label{superfluid-eq10}\eea
This assumes that
$\hbar\omega_0 \ll k_BT\ ; \ gn_c (\br, t) \ll k_BT$, namely both the 
energy level spacing in the harmonic well trap and the average mean field 
should be less than the average kinetic energy of the atoms.  Under these 
conditions, the high-energy thermal atoms can be described by the 
single-particle distribution function $f({\bf p}, \br, t)$ obeying a 
Boltzmann equation
\begin{eqnarray}
{\partial f({\bf p}, {\bf r}, t)\over\partial t} &+&{{\bf p}\over 
m}\cdot\mbox{\boldmath$\nabla$}_r  
f({\bf p}, {\bf r}, t) - \mbox{\boldmath$\nabla$}_r 
U({\bf r}, t) \cdot\mbox{\boldmath$\nabla$}_p f({\bf p}, {\bf r}, t) 
\nonumber \\
&=& C_{22}[f] + C_{12}[f, \Phi].\label{superfluid-eq12}\end{eqnarray}
This involves two types of collision integrals.  Collisions between 
non-condensate atoms are described by \begin{eqnarray}
C_{22}[f] &=& {2g^2\over (2\pi)^5}\int d{\bf p}_2\int d{\bf p}_3\int d{\bf 
p}_4\delta({\bf p}+{\bf p}_2 - {\bf p}_3 - {\bf p}_4)\nonumber \\
&\times&\delta\left({\tilde\varepsilon}_p+
{\tilde\varepsilon}_{p_2}-{\tilde\varepsilon}_{p_3}
-{\tilde\varepsilon}_{p_4}\right)\nonumber 
\\
&\times&\left[(1+f)(1+f_2)f_3f_4 - 
ff_2(1+f_3)(1+f_4)\right].\label{superfluid-eq13}\end{eqnarray}
In contrast, collisions which transfer atoms between the condensate and 
non-condensate are described by
\begin{eqnarray}
C_{12}[f, \Phi] &=& {2g^2\over (2\pi)^2}\int d{\bf p}_1 \int d{\bf p}_2 
\int d{\bf p}_3\delta (m{\bf v}_s +{\bf p}_1-{\bf p}_2-{\bf p}_3) 
\nonumber\\
&\times& \delta\left(\varepsilon_c 
+{\tilde\varepsilon}_{p1}-{\tilde\varepsilon}_{p2}
-{\tilde\varepsilon}_{p3}\right)\left[\delta\left({\bf 
p}-{\bf p}_1)-\delta({\bf p}-{\bf p}_2)-\delta({\bf p}-{\bf 
p}_3\right)\right]\nonumber \\
&\times&\left[n_c(1+f_1)f_2f_3 
-n_cf_1(1+f_2)(1+f_3)\right].\label{dynamics-eq14}\end{eqnarray}
Here $\varepsilon_c(\br, t) =\mu_c +{1\over 2}mv^2_s$ is the condensate 
atom local energy $(\mu_c$ is defined below in (\ref{superfluid-eq18})) 
and $m{\bf v}_s$ is the condensate atom momentum.

Clearly the $C_{12}$ collisions do not conserve the number of atoms in the 
condensate.  Thus they enter (in contrast to $C_{22}$ collisions) in a 
direct way in a ``generalized'' GP equation for $\Phi(\br, t)$, 
namely\cite{Zarnikgrif,Grifsum}
\begin{equation}
i\hbar{\partial\Phi\over\partial t} = \left[-{\hbar^2\nabla^2\over 
2m}+V_{ex} ({\bf r}) + gn_c({\bf r}, t) + 2g{\tilde n}({\bf r}, t) -i\hbar 
R({\bf r}, t)\right]\Phi\ .\label{superfluid-eq15}\end{equation}
There is a new term $(2g{\tilde n})$ associated with the Hartree-Fock mean 
field of the thermal cloud on the condensate.  In addition, there is a 
dissipative term associated with $C_{12}$ collisions, 
\begin{equation}
R({\bf r}, t) =\int{d{\bf p}\over(2\pi)^3} {C_{12}[f, \Phi]\over 2n_c(\br, 
t)}\equiv{\Gamma_{12}[f, \Phi]\over 2n_c({\bf r}, t)} \sim 
0(g^2).\label{superfluid-eq16}\end{equation}
As with the $T=0$ GP equation of Section~\ref{sec:Dynamics}, one can 
rewrite (\ref{superfluid-eq15}) in terms of the $n_c(\br, t)$ and 
${\bv}_s(\br, t)$ variables to find (compare with (\ref{dynamics-eq6}) 
and  (\ref{dynamics-eq7}))
\bea
{\partial n_c\over\partial t}+\mbox{\boldmath$\nabla$}\cdot n_c{\bf v}_c 
&=&-\Gamma_{12}[f, \Phi]\nonumber\\
m\left({\partial{\bf v}_c\over\partial t} + {1\over 
2}\mbox{\boldmath$\nabla$}{\bf v}_c^2\right) &=& 
-\mbox{\boldmath$\nabla$}\mu_c\ ,\label{superfluid-eq17}\eea
where now
\begin{equation}
\mu_c({\bf r}, t) = -{\nabla^2\sqrt{n_c}\over 2m\sqrt{n_c}} + V_{ex}({\bf 
r}) + gn_c({\bf r}, t) + 2g{\tilde n}({\bf r}, t).\label{superfluid-eq18}
\end{equation}
We note that even at finite $T$, the condensate equations of motion in 
(\ref{superfluid-eq17}) are ``hydrodynamic looking'' except that now in 
the continuity equation for the condensate, there is a source term 
$\Gamma_{12}$ associated with $C_{12}$ collisions.  

 One has to solve these coupled equations for $\Phi(\br, t)$ and  $f({\bf 
p}, \br, t)$ self-consistently - and clearly $C_{12}$ will play a special 
role.  While (\ref{dynamics-eq4}) and 
(\ref{superfluid-eq15}) give a sound basis for discussing the general 
non-equilibrium behavior of trapped gases at finite $T$, the 
collision-dominated hydrodynamic region is especially interesting since 
then the non-condensate can be described in terms of a few 
``coarse-grained'' local variables. If the $C_{22}$ collisions are rapid 
enough (relative to the frequency $\omega$ of the collective mode of 
interest, $\omega\tau_{22}\ll 1)$, then $f({\bf p}, \br, t)$ will be 
driven to the {\it local} equilibrium Bose distribution
\begin{equation}
{\tilde f}({\bf p}, {\bf r}, t) = {1\over e^{\beta[{({\bf p}-m{\bf 
v}_n)^2\over 2m}+U({\bf r}, t) - {\tilde\mu}(r, t)]}-1}. 
\label{superfluid-eq19}\end{equation}
This function uniquely satisfies $C_{22}[{\tilde f},\Phi]=0.$
With this local equilibrium distribution, the thermal atoms are now 
completely described in terms of the $\beta, {\bf v}_n, {\tilde\mu}$ and 
${\tilde U}$ variables, all dependent on $(\br, t)$.  Inserting ${\tilde 
f}$ given by (\ref{superfluid-eq19}) into the Boltzmann equation 
(\ref{dynamics-eq4}) and taking moments with respect to the momentum ${\bf 
p}$, one easily derives a set of hydrodynamic equations for the 
non-condensate variables ${\tilde n}, {\bf v}_n$ and ${\tilde\mu}$.  This 
procedure was first developed by Boltzmann for deriving the hydrodynamic 
equations for a classical gas.

To summarize, at this stage we have arrived at a closed set of coupled 
equations\cite{Zarnikgrif,Nikuni99,Grifsum} for the condensate variables 
$(n_c, {\bf v}_s, \mu_c)$ and the non-condensate variables $({\tilde n}, 
{\bf v}_n, {\tilde\mu})$.  Linearizing these equations around static 
equilibrium, one can derive the hydrodynamic collective modes which  
correspond to coherent motions of both the condensate and non-condensate.  
Our new two-fluid equations exhibit a new relaxation time $\tau_\mu,$ 
which describes how fast the two components  come into diffusive 
equilibrium with each other (i.e., how fast $\mu_{diff}(\br, t) 
\equiv{\tilde\mu}(\br, t) -\mu_c(\br, t)$ relaxes to zero). Of course, in 
static equilibrium, we have ${\tilde\mu}_0 =\mu_{c0}$ but when we perturb 
the trapped Bose gas from equilibrium, one can have $\delta\mu_{diff}(\br, 
t)\ne 0$ and it relaxes to zero in a time $\tau_\mu.$  As one would 
expect, $\tau_\mu$ is proportional to the collision time $\tau_{12}$ 
associated with $C_{12}$ collisions.

How is all this related to Landau's famous two-fluid hydrodynamic 
equations?  These equations are expressed using different local 
variables.  The linearized form of these two-fluid equations 
are\cite{Khalatnikov}
\bea
{\partial\delta n\over\partial t} &=& 
-\mbox{\boldmath$\nabla$}\cdot\delta{\bf j}\nonumber\\
 m{\partial\delta j_\mu\over\partial t}&=& -{\partial\delta P\over\partial 
x_\mu}-\partial n{\partial V_{ex}\over\partial 
x_\mu}+{\partial\over\partial x_\nu}\left[2\eta\left(D_{\mu\nu}-{1\over 
3}TrD\delta_{\mu\nu}\right)\right]\nonumber\\
{}&+&{\partial\over\partial x_\mu}\left[\zeta_1
\mbox{\boldmath$\nabla$}\cdot(mn_{c0}(\delta{\bf v}_s-\delta{\bf 
v}_n)+\zeta_2\mbox{\boldmath$\nabla$}\cdot\delta{\bf v}_n\right]\nonumber\\
m{\partial\delta {\bf v}_s\over\partial t} &=& 
-\mbox{\boldmath$\nabla$}\left[\delta\mu
+m\zeta_3\mbox{\boldmath$\nabla$}\cdot mn_{c0}(\delta{\bf v}_s-\delta{\bf 
v}_n)+m\zeta_4\mbox{\boldmath$\nabla$}\cdot\delta{\bf 
v}_n\right]\nonumber\\
{\partial\delta s\over\partial t} 
&=&-\mbox{\boldmath$\nabla$}\cdot(s_0\delta{\bf v}_n)+{1\over 
T}\mbox{\boldmath$\nabla$}\cdot(\kappa\mbox{\boldmath$\nabla$}\delta T) 
.\label{superfluid-eq20}\eea
Here $P(\br, t)$ is the local pressure, $s(\br, t)$ is the local entropy 
density (entirely associated with the normal fluid), and 
$D_{\mu\nu}\equiv{1\over 2}\left({\partial v_{n\mu}\over\partial 
x_\nu}+{\partial v_{n\nu}\over\partial x_\mu}\right).$   The expressions
\bea\delta \rho&\equiv& m \delta n = \delta\rho_s+\delta\rho_n\nonumber\\
m\delta{\bf j} &\equiv& \rho_{s0}\,\delta\bv_s 
+\rho_{n0}\,\delta\bv_n.\label{superfluid-eq21}\eea
make the two-fluid nature of the theory clear.  We have included 
hydrodynamic damping from various transport processes (the thermal 
conductivity $\kappa$, the shear viscosity $\eta$ and the four second 
viscosity coefficients $\zeta_i$, all dependent on position through the 
local condensate density $n_{c0}(\br)$). In making the detailed comparison 
between the Landau two-fluid equations and our two-fluid equations, a key 
role is played by the condensate relaxation time $\tau_\mu$  introduced 
above.

In the limit $\omega\tau_\mu\to 0,$ one can show\cite{Zarnikgrif} that our 
two-fluid equations are precisely equivalent to the Landau equations 
summarized in (\ref{superfluid-eq20}), but without the dissipative terms. 
This makes sense, since these equations are derived\cite{Bogg} under the 
assumption that the superfluid and normal fluid are always in local 
equilibrium with each other.  This corresponds to ${\tilde\mu}(\br, t) 
=\mu_c(\br, t),$ ie, $\tau_\mu\to 0.$  However, even in this limit, it 
turns out that $\Gamma_{12}$ in (\ref{superfluid-eq17}) is finite and is 
crucial to ensure the equivalence with the Landau equations.  One finds 
one can make the identification (valid within our model)
\bea \rho_s(\br, t) &=& mn_c(\br, t)\nonumber\\
\rho_n(\br, t) &=& m{\tilde n}(\br, t).\label{superfluid-eq22}\eea
 
In the limit of $\omega\tau_\mu\ll 1, \omega\tau_{22} \ll 1$, our 
equations can be shown\cite{Tnik} to reduce precisely to the 
Landau-Khalatnikov two-fluid equations\cite{Khalatnikov} in 
(\ref{superfluid-eq20}).  What is very satisfactory is that, in our model, 
the second viscosity damping terms turn out to be proportional to 
$\delta\mu_{diff}$, i.e., to the fact that $\mu_c\ne{\tilde\mu}.$  This is 
physically quite reasonable and expected.  Thus the four second 
viscosities are all proportional to the collision time $\tau_{12}$ 
associated with the $C_{12}$ collision term.  In contrast, the other 
transport coefficients $(\kappa$ and $\eta$) are associated with 
deviations for $f({\bf p}, \br, t)$ from the local equilibrium 
distribution  (\ref{superfluid-eq19}) and have contributions from {\it 
both} the $C_{22}$ and $C_{12}$ collision integrals.

Finally, one can consider the new limit $\omega\tau_\mu\gg 1, 
\omega\tau_{22}\ll 1,$ which can arise\cite{Nikuni99} at temperatures 
close to $T_{BEC}$ (where $\tau_\mu$ becomes very large due to a sort of 
``critical slowing down'').  This is outside the region of validity of the 
usual Landau two-fluid equations.  Surprisingly, however, our new 
linearized two-fluid equations lead again\cite{Tnik} to the 
Landau-Khalatnikov two-fluid equations\cite{Khalatnikov} but now with 
frequency-dependent second viscosities

\beq \zeta_i(\omega) = {\zeta_i\over 
1-i\omega\tau_\mu}.\label{superfluid-eq23}\eeq
This form could have been predicted, since it is typical of a situation 
where a fluid is coupled into an ``internal'' degree of freedom with a 
long relaxation time - in the present case, this is the Bose condensate.

As one might expect, our new form of the two-fluid equation leads 
naturally to a central zero-frequency relaxational mode which is missed by 
the usual LK two-fluid hydrodynamic equations.  One can 
show\cite{Tnikgrifezar,Grifsum} that as $T\to T_{BEC}$ from below, this 
mode (which is strongly coupled to the thermal conductivity) becomes the 
usual zero frequency thermal diffusion mode.  Observation of this mode 
below $T_{BEC}$ is a goal for future BEC studies since this mode is a 
unique feature of the hydrodynamics of a trapped Bose gas. 

In conclusion, we have sketched recent progress in deriving the equivalent 
of the dissipative Landau-Khalatnikov two-fluid hydrodynamic equations for 
a trapped Bose-condensed gas.  The nice feature about our derivation is 
that our microscopic model allows us to compute all the thermodynamic 
functions, transport coefficents and relaxation times which enter into 
these equations.  What is needed now is a sustained experimental effort to 
check the predictions of this two-fluid theory.
\section*{ACKNOWLEDGMENTS}
\label{sec:ack}
Much of the work presented in Section 3 has been done in close 
collaboration with  Eugene Zaremba.  A.G. is supported by NSERC of Canada 
and T.N. is supported by the JSPS.

\end{document}